\documentclass[%
 aip,
 jmp,%
 amsmath,amssymb,
 reprint,%
]{revtex4-1}

\usepackage{graphicx}
\usepackage{dcolumn}
\usepackage{bm}

\begin{document}

\preprint{AIP/123-QED}

\title{A novel vacuum ultra violet lamp for metastable rare gas experiments}

\author{Heiner Daerr}
\author{Markus Kohler}
\author{Peter Sahling}
\author{Sandra Tippenhauer}
\author{Ariyan Arabi-Hashemi}
\affiliation{Carl Friedrich von Weizs\"cker Centre for Science and Peace Research (ZNF), University of Hamburg, 20144 Hamburg, Germany}

\author{Christoph Becker}
\author{Klaus Sengstock} 
\affiliation{Institut f\"ur Laser-Physik, University of Hamburg, 22761 Hamburg, Germany}

\author{Martin B. Kalinowski}
\affiliation{Carl Friedrich von Weizs\"acker Centre for Science and Peace Research (ZNF), University of Hamburg, 20144 Hamburg, Germany}

\date{\today}

\begin{abstract}
We report on a new design of a vacuum ultra violet (VUV) lamp for direct optical excitation of high laying atomic states e.g. for excitation of metastable rare gas atoms. 
The lamp can be directly mounted to ultra high vacuum vessels (p$\,\leq\,$10$^{-10}\,$mbar).  
It is driven by a 2.45$\,$GHz microwave source.
For optimum operation it requires powers of approximately 20$\,$W.
The VUV light is transmitted through a magnesium fluoride window, which is known to have a decreasing transmittance for VUV photons with time.
In our special setup, after a run-time of the VUV lamp of 550$\,$h the detected signal continuously decreased to 25$\,$\% of its initial value.
This corresponds to a lifetime increase of two orders of magnitude  compared to previous setups or commercial lamps. 
\end{abstract}

\maketitle


\section{Introduction}

Laser cooled metastable rare gas atoms are of high interest in cold collision physics \cite{Walhout1995}, optical lattices \cite{Kunugita1997}, atom lithography \cite{Berggren1995}, Bose-Einstein condensation \cite{Robert2001} and atom trap trace analysis (ATTA) \cite{Chen1999}. 
A common requirement of the aforementioned experiments is the need for a reliable high-flux source of metastable atoms. \\
We are currently setting up a new highly efficient apparatus for detecting Kr-85 and Kr-81 isotopes in small environmental samples.
ATTA is considered to be the most promising technique to measure such long-lived radioactive rare gas isotopes \cite{Collon2004,Lu2010}.
The atom of interest is trapped inside a magneto-optical trap (MOT) and detected by the fluorescence originating from the trapping process.
The main challenge for an efficient ATTA system for krypton isotopes is the production of the metastable state (5s[3/2]$_2$) which is trappable by optical means.\\
Typically, metastable krypton atoms for trace analysis are produced by a RF discharge \cite{Chen2001}, which has some profound disadvantages like a low efficiency of Kr*/Kr$\,$$\approx 10^{-4}$, an increased atomic beam divergence, the unavoidable production of krypton ions causing cross contaminations and very importantly the requirement of a minimum gas pressure to operate, which sets a lower limit on the sample size.\\ 
The production of metastable atoms by a purely optical excitation scheme \cite{Young2002} overcomes the limitations originating from the RF discharge.
In practice, optical excitation has not been applicable so far, due to the short lifetime of VUV lamps of a few hours \cite{Okabe1964, Lu2010}. \\
We have developed a VUV lamp for the efficient and reliable production of metastable krypton (Kr*) with an extended lifetime of hundreds of hours making optical excitation implementable in atom trap trace analysis and other applications. 
We carefully analyzed important lamp parameters such as the transmittance of the magnesium fluoride window over time.\\ 
The capability of measuring small sample sizes will enable groundbreaking  new applications in the fields of environmental and earth sciences \cite{Corcho2007, Sturchio2004, Bauer2001} as well as nuclear non-proliferation \cite{Daerr2010}.

\section{VUV lamp design} \label{VUVLmap}
\begin{figure}[h]
	\centering
		\includegraphics{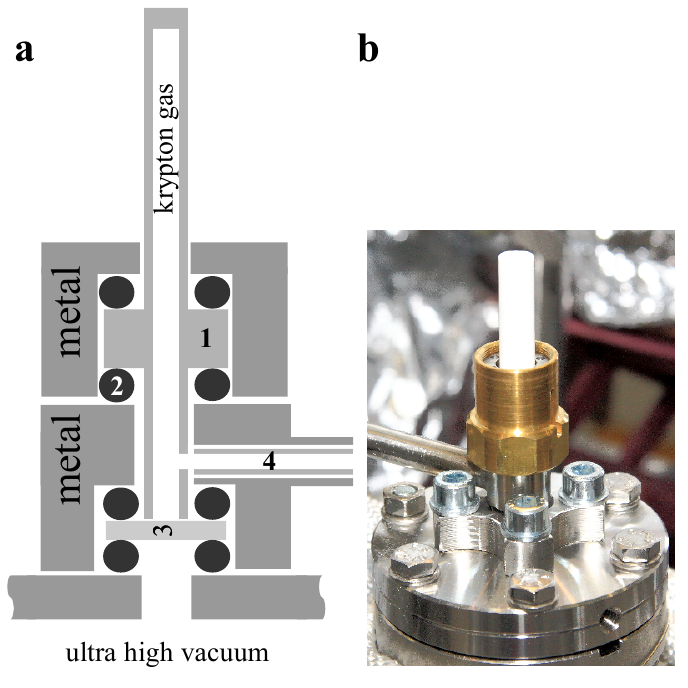}
	\caption{Figure a shows a sketch of the VUV lamp. 
The gas discharge producing the VUV photons takes place inside a hollow Macor rod (part~1) at krypton pressures of $\approx 2\,\mathrm{mbar}$. 
The rod is sealed using two resilient metal seals (part~2). 
The light is emitted through a magnesium fluoride window (part~3), which is again sealed with resilient metal seals.
The lamp is evacuated to pressures below $10^{-9}\,$mbar and subsequently filled with pure krypton using one tube (part~4), which is lined with Macor to reduce the contact of plasma and metal.
Figure \textbf{b} shows \textbf{a} photograph of a lamp with the surfatron dismounted.}
	\label{fig:VUVLamp}
\end{figure}

We designed a VUV lamp with regard to an extended lifetime.
The lifetime is limited because the transmittance of the magnesium fluoride window in the VUV spectral region decreases with run-time \cite{Okabe1964, Lu2010}.
The degeneration might be explained by several effects like formation of color centers (F-center) \cite{Saito1996, Facey1969, Blunt1967} or adsorption  of water, molecular oxygen, methane or non-methane hydrocarbons at the surface or the bulk of the magnesium fluoride window \cite{Koch1965, Herzig1992}.
We  focused on the reduction of the last mentioned pollutions inside the lamp successfully increasing the lifetime by using resilient metal seals and continuously purifying the krypton gas.\\
Figure  \ref{fig:VUVLamp} shows a sketch of the new VUV lamp design.
The microwave driven discharge generating the VUV photons takes place inside a hollow Macor rod filled with pure krypton gas.
The VUV photons are transmitted through a magnesium fluoride window sealed with resilient metal seals.
It can be flanged to ultra high vacuum chambers. 
The gas inlet of the lamp is attached to a titan sublimation pump filled with krypton acting as a reservoir to continuously change the gas inside the lamp and to maintain a high vacuum environment beside the krypton partial pressure. 
Krypton gas with a purity of 4.7, which was further purified using a gas purifier, was used.
The inlet is lined with Macor to reduce the deposition of material on the window, which was removed from the metallic vacuum components by the plasma. \\
A home-made surfatron \cite{Moisan1979, Moisan1991} operating at 2.45$\,$GHz is used to run the discharge at powers of $20\,$W.
The resident heating is compensated by water cooling.
A surfatron was chosen as it produces long plasma columns, so the discharge extends to the window surface reducing self-absorption \cite{Cowan1948}.
The compact lamp design allows to place many VUV lamps close to each other with a clearance of 25$\,$mm from center to center to increase the total VUV photon flux.

\section{Performance of the VUV lamp}
\begin{figure*}
	\centering
		\includegraphics{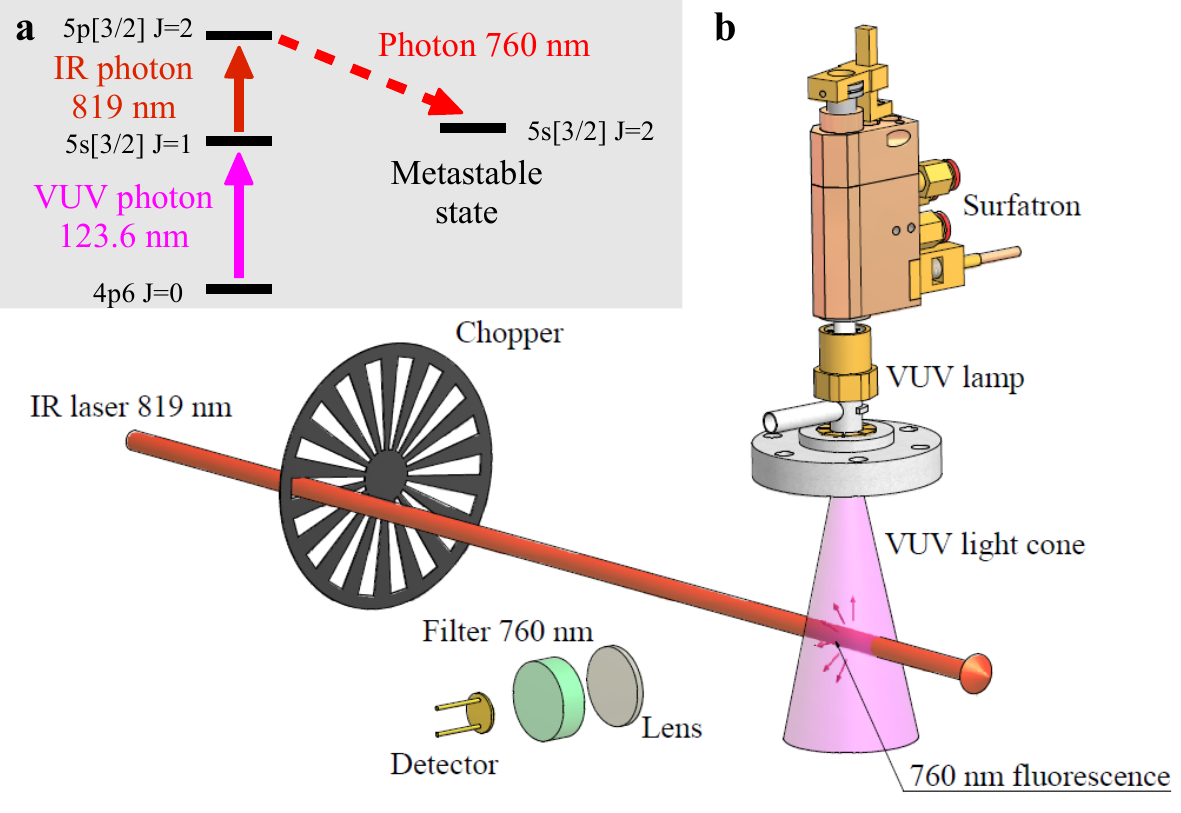}
	\caption{Optical excitation scheme of krypton (Figure a) and a sketch of the experimental setup (Figure b).
The excitation scheme is a resonant two-photon excitation from the ground state 4p6  to the 5p[3/2]$_2$ state which can spontaneously decay to the metastable state 5s[3/2]$_2$ by emitting a photon at 760$\,$nm.  
Three electronic dipole transitions need to be involved as the ground level has an even parity and an angular momentum $J=0$ and the metastable state has an odd parity and $J=2$.
The fluorescence at 760$\,$nm is a measure for production of the metastable state.
For the detection of fluorescence the laser light was modulated with an optical chopper to apply lock-in detection techniques.
The fluorescence light was focused on a self-made photo detector by a lens system with an optical filter at 760$\,$nm.}
	\label{fig:AbbExp}
\end{figure*}

\begin{figure}
	\centering
		\includegraphics{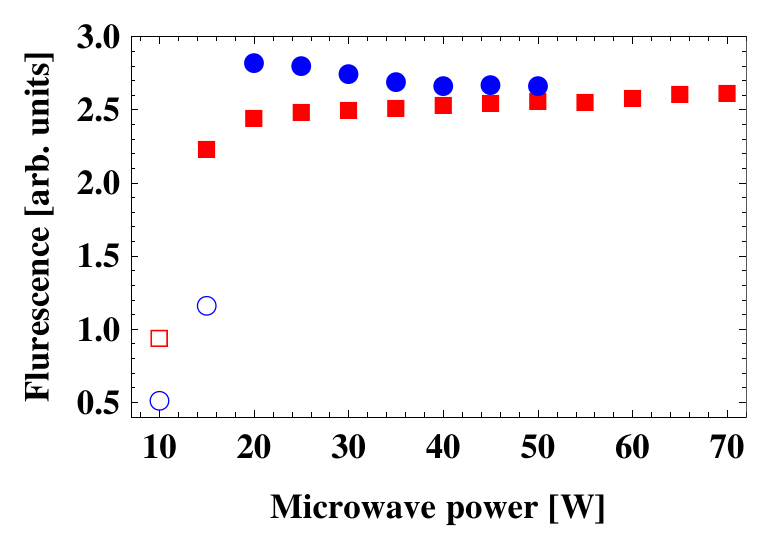}	
	\caption{The Figure shows the fluorescence at 760$\,$nm as a function of the microwave power for two different krypton pressures 	2.0$\,$mbar (filled points) and 3.1$\,$mbar (filled squares) inside the VUV lamp. 
A pressure of 2.0$\,$mbar inside the lamp produces a slightly higher fluorescence with a maximum at 20$\,$W microwave power in comparison to the fluorescence at 3.1$\,$mbar. 
The constant increase of fluorescence with increasing microwave power is small for a pressure of 3.1$\,$mbar. 
The data points at 10$\,$W for both pressures and the data point at 15$\,$W for 2.0$\,$mbar belong to unstable operation conditions of the VUV lamp (open points and squares).}
	\label{fig:Power}
\end{figure}

\begin{figure}
	\centering
		\includegraphics{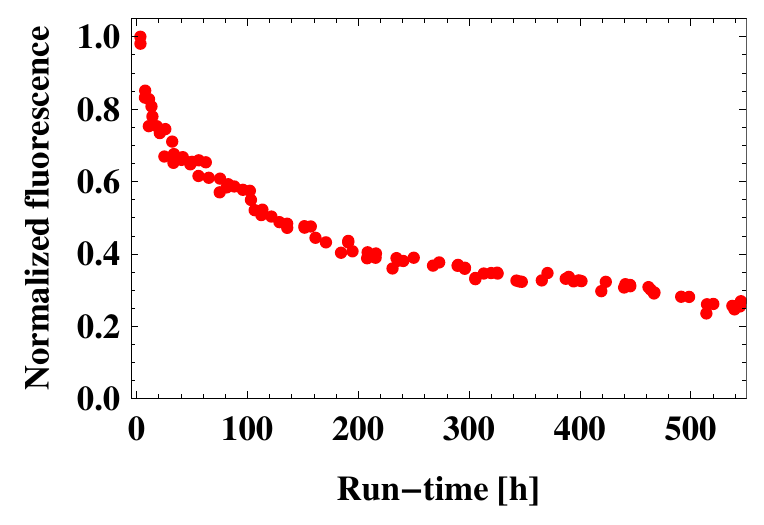}
	\caption{The fluorescence at 760$\,$nm was recorded for a duration of 550$\,$h under the same conditions (20$\,$W microwave power, 2.0$\,$mbar lamp pressure, 0.5$\,$mbar pressure in detection chamber, 10$\,$mW laser power).
	The fluorescence signal decreased to 25$\,$\% of its initial value during measurement.}	
	\label{fig:LampeLangzeit}
\end{figure}

Within the setup depicted in Figure \ref{fig:AbbExp} the lamp was successfully used to excite krypton atoms to the metastable state.
An estimated production rate of 10$^9$$\,$-$\,$10$^{10}$ krypton atoms/s was achieved in a small excitation volume 100$\,$mm below the lamp at a pressure of about 0.5$\,$mbar krypton inside the detection chamber. 
Note: The production rate strongly depends on the experimental geometry and is in the current setup limited due to the large distance between the excitation volume and the magnesium fluoride window. 
We expect a much larger excitation rate directly below the window.\\
In this setup an exact assessment of the resulting VUV photon flux of the lamp from the detected production rate is difficult due to the unknown spectral and spatial distribution of the VUV photons, the exact imaged excitation volume and the unknown fraction of VUV photons reaching the excitation volume.\\
In the following we describe the operation and performances of the VUV lamp with focus on the particularly important aspect of the lifetime.
To explore the performance and the temporal behavior of the power emitted by the lamp, krypton gas was excited to the metastable state using the VUV lamp in combination with a laser stabilized to the 819$\,$nm transition in Kr-84 (compare Figure \ref{fig:AbbExp}).\\
The VUV lamp operates at krypton pressures from 0.9$\,$mbar to 5$\,$mbar.
Microwave powers of about 20$\,$W are sufficient to maintain a stable discharge.
Beyond a certain threshold, neither the pressure inside the lamp nor the microwave power have strong influence on the strength of the fluorescence at 760$\,$nm over a wide range (see Figure \ref{fig:Power}). 
The fluorescence is a direct measure of the production rate of metastable atoms.\\
We know turn to the important question of the lifetime of the VUV lamp.  
Therefore the  fluorescence at 760$\,$nm was detected over time as a measure for the transmittance of the magnesium fluoride window, as the production of metastable atoms depends linearly on the VUV intensity for our experimental parameters.
For the assessment of the lifetime of the lamp the fluorescence of a fresh lamp was monitored for 550$\,$h under the same conditions.
The gas was renewed twice a day.
After 550$\,$h run-time the fluorescence signal decreased to 25$\,$\% of the initial value (see figure \ref{fig:LampeLangzeit}).
This corresponds to a lifetime increase of two orders of magnitude  compared to previous setups or commercial lamps.\\
Note that the data for Figure \ref{fig:Power} was recorded with another lamp than the one used to collect the data shown in Figure \ref{fig:LampeLangzeit}.  
After recording the data for Figure \ref{fig:Power} the measured fluorescence rapidly drops almost to zero.
This rapid decrease of the transmittance indicates, that the degeneration of the window heavily depends on the microwave power.
Therefor it is important to operate the VUV lamp at a microwave power as low as possible to extend the lifetime of the VUV lamp. 
\section{Conclusion and Outlook}
We have developed a new design of a VUV lamp with an extended lifetime.
The lifetime of the VUV lamp is about two orders of magnitude greater compared to commercially available lamps.
This makes the all optical excitation scheme for metastable rare gas atoms applicable in state-of-the-art experiments e.g. for atom trap trace analysis whose efficiency is seriously affected by the poor performance of the RF discharge excitation.\\ 
The lamp can be attached to ultra high vacuum environments required for efficient trapping of atoms in a MOT.
Contamination of the vacuum system due to poor sealing of the lamp is avoided.
To overcome  the power limitation of VUV light, a small sized lamp was developed for the purposes of mounting many lamps with a small clearance (25$\,$mm).\\

\begin{acknowledgments}

We gratefully acknowledge the technical support of the staff members of the mechanic workshop of the department of chemistry  of the University of Hamburg in developing a small sized VUV lamp. 
Further more we thank Frank Buerli for supplying vacuum components and the microwave generator.
We thank Franziska Herrmann for calculating the imaging optics.\\
We thank the Deutsche Forschungsgemeinschaft and the German Foundation for Peace Research for funding.

\end{acknowledgments}

\end{document}